\documentclass[12pt,preprint]{aastex}

\begin{document}
\title{Late-Time Convection in the Collapse of a 23 Solar Mass Star}

\author{Christopher L. Fryer\altaffilmark{1,2} and Patrick
A. Young\altaffilmark{2,3}}

\altaffiltext{1}{Department of Physics, The University of Arizona,
Tucson, AZ 85721} 
\altaffiltext{2}{Theoretical Division, LANL, Los Alamos, NM 87545}
\altaffiltext{3}{Steward Observatory, The University of Arizona,
Tucson, AZ 85721} 

\begin{abstract}

The results of a 3-dimensional SNSPH simulation of the core collapse
of a 23\,M$_\odot$ star are presented.  This simulation did not launch
an explosion until over 600\,ms after collapse, allowing an ideal
opportunity to study the evolution and structure of the convection
below the accretion shock to late times.  This late-time convection
allows us to study several of the recent claims in the literature
about the role of convection: is it dominated by an $l=1$ mode driven
by vortical-acoustic (or other) instability, does it produce strong
neutron star kicks, and, finally, is it the key to a new explosion
mechanism?  The convective region buffets the neutron star, imparting
a $150-200 {\rm \, km \, s^{-1}}$ kick.  Because the $l=1$ mode does not
dominate the convection, the neutron star does not achieve large
($>450 {\rm \, km \, s^{-1}}$) velocities.  Finally, the neutron star in
this simulation moves, but does not develop strong oscillations, the
energy source for a recently proposed supernova engine.  We discuss
the implications these results have on supernovae, hypernovae (and
gamma-ray bursts), and stellar-massed black holes.

\end{abstract}


\section{Introduction}

Since Epstein (1979) first proposed that convection could increase the
neutrino luminosity arising from newly formed neutron star,
core-collapse theorists have studied the potential roles convection
may play in the supernova explosion mechanism.  This work has
identified 2 different regions where instabilities are produced: (1)
within the proto-neutron star and (2) the region between the accretion
shock and the outer surface of the proto-neutron star (Fig. 1).
Convection within the proto-neutron star originates initially from the
lepton gradients near the neutrinosphere as proposed by Epstein, but
the convective instabilities that grow involve much more complex
criteria than simple weight gradients (Keil et al. 1996, Bruenn et
al. 2004).  If the convection can grow, it can help transport
neutrinos out of the core, effectively increasing the luminosity of
neutrinos at early times.  The strength of this convection is still an
important topic of study and it remains to be seen if this convection
grows beyond the limited region near the neutrinosphere: indeed, most
simulations (either rightly or wrongly) do not exhibit strong
proto-neutron star convection (e.g. Herant et al. 1994; Fryer \&
Warren 2002; Buras et al. 2003).  The role of this convection can only
be answered with a detailed knowledge of the equation of state at
nuclear densities (Bruenn et al. 2004).

On the other hand, convection above the proto-neutron star tends to be
strong in most multi-dimensional models (e.g. Herant et al. 1994;
Burrows et al. 1995; Mezzacappa et al. 1998; Fryer \& Warren 2002;
Buras et al. 2003; Burrows et al. 2006).  This convection, initially
driven by the entropy gradient left behind after the stall of the
bounce shock aids the conversion of heat released by gravitational
potential energy during the infall into kinetic energy of an explosion
(see Fryer 2003 for details).  One of the debates with this convection
is the role it plays in helping to drive an explosion.  Herant et
al. (1994) focused on the role convection played in transporting the
neutrino-heating material out of the star allowing this thermal energy
to convert to kinetic energy.  The convective region is also a
resevoir for the energy ``useful'' energy, that is, energy that can
drive an explosion.  But until recently, the effect of this convection
met with mixed results where some multi-dimensional simulations
achieved strong explosions due to convection and others produced no
explosion whatsoever (compare Herant et al. 1994 to Mezzacappa et
al. 1998).  But as we shall discuss below, there is a growing
consensus that this convective region is important in core-collapse
supernovae.  Indeed, convection has been recently proposed as not just
an aid in converting neutrino heat energy into an explosion, but as
the primary means of generating the energy behind the supernova
explosion (Burrows et al. 2006).

Another growing debate surrounding the convective region above the
proto-neutron star focuses on the cause of this instability.  Herant
et al. (1994) originally focused on the entropy gradient set by the
stall of the bounce shock and the entropy generated by neutrino
heating at the base of this convective region.  Shock heating produced
by the downflows as they strike the hard surface of the proto-neutron
star also contributes to the convective instability of the region.
Blondin et al. (2003) focused their analysis on the instabilities
produced at the top of this convective region, also termed accretion
shock instabilities.  Due to the low-mode nature of their convection,
they argued for a new (third) type of instability extracting the
energy stored in vorticity and converting this energy into sound
waves, the ``voritical-acoustic'' instability.  Although this
instability has attracted the attention of additional supernova groups
(e.g. Burrows et al. 2006), Blondin \& Mezzacappa (2006) remain
prudently wary (if not doubtful) of their own proposal, focusing
instead on the more classical drivers behind accretion shock
instabilities (Houck \& Chevalier 1992).  But Burrows et al. (2006)
have found that this vortical-acoustic instabilities drove
oscillations in the proto-neutron star that ultimately can power a
supernova explosion.

Convection above the proto-neutron star may also be the cause of the
strong kicks observed in the pulsar population and required to explain
a number of binary systems (see Fryer et al. 1998 and Lai et al. 2001
for reviews).  Herant (1995) used the simulations of Herant et
al. (1994) to argue that if such modes merged, a strong kick could be
produced.  Scheck et al. (2004) have now run a series of simulations
showing that such a kick mechanism is not only plausible, but 
can also produce the obvserved pulsar velocity distribution.

It is difficult to compare/contrast this large set of
multi-dimensional core-collapse calculations.  Core-collapse is a
complex problem, with a wide range of physics that can play an
important role in the development of the explosion.  Different
calculations have different implementations (with varying levels of
sophistication) of this physics.  Many of these results are based on
2-dimensional simulations with simplified treatments of the neutron
star.  Others do not model convection out to late times, an important
factor in the development of low mode convection.  In this paper, we
present the results of the collapse of a 23\,M$_\odot$ star, following
the convection $\sim$400\,ms after bounce (over 600\,ms after the
collapse of the massive star).  In \S 2, we discuss the code used in
these calculations and its strengths and weaknesses versus other
techniques.  \S 3 focuses on the convective region above the
proto-neutron star and the evolution of this convection with an effort
to distinguish between numerical and real results.  \S 4 moves this
study inward to the motion and evolution of the proto-neutron star.
We conclude with a discussion of the implications of these results on
our current understanding of supernovae.

\section{Initial Conditions and Numerical Techniques}

Our initial progenitor is a 23\,M$_\odot$ star produced by the Tycho
stellar evolution code (Young \& Arnett 2005).  The Tycho code itself
is evolving away from the classic technique (mixing-length theory) of
modeling convection to a more realistic algorithm based on
multi-dimensional studies of convection in the progenitor star (Meakin
et al. 2005).  In Figure 2 we see that the density and entropy
structure of this progenitor is quite different than those produced by
classic stellar evolution codes such as Kepler (Heger et al. 2006).
One key difference is the lack of jumps in the density and temperature
profile in the star produced using the Tycho stellar evolution code.
The jumps are an artifact of mixing length theory, which ignores
hydrodynamic tranport processes at and outside of the convective
boundary. In shell burning especially, inner convective boundaries are
stiff, but outer boundaries are soft. Due to the large bouyancy
frequencies at the boundary, relatively low mach number flows ($M \sim
0.01$) can generate waves in the intershell regions with density
contrasts of $\delta \rho / \rho$ of order 10\% in oxygen burning
(Meakin \& Arnett 2006b). These processes smooth the temperature and
density gradients. More realistic treatment of the convective and
boundary hydrodynamics also results in larger convective zones.
The differences in the structure will affect the fate of the star, and
it is likely that uncertainties in the progenitor dominate the
uncertainties in any core-collapse calculation.  We also compare this
structure to the structure of the classic 15\,M$_\odot$ progenitor
(s15s7b2: Woosley \& Weaver 1995) used as a standard in many
core-collapse calculations.  Note that higher mass stars have cores
with higher entropies.  The 23\,M$_\odot$ progenitor used in this
study is produced with a version of the Tycho code part way through
this transformation, and we expect the exact structure of this
23\,M$_\odot$ to change as stellar evolution codes improve (although
preliminary results suggest the changes with the fully transformed
code will be small  - Young, pvt. communication).

The higher densities in the 23\,M$_\odot$ models over the
15\,M$_\odot$ star lead to higher accretion rates (and a higher ram
pressure) at the top of the convective region.  It is this infalling
material that prevents the supernova explosion (Fryer 1999).  Figure 3
shows the accretion rates for the 3 progenitors: the 23\,M$_\odot$ in
this study from Tycho (Young \& Fryer 2006), a 23\,M$_\odot$ Kepler
model (Heger et al. 2006), and the 15\,M$_\odot$ standard model
(Woosley \& Weaver 1995).  The 23\,M$_\odot$ models have a higher
accretion rate, and hence higher ram pressure to overcome to drive a
supernova explosion.  Here is where the difference between stellar
evolution codes truly stands out.  Note the large difference between
accretion rates.  The difference between the Kepler and the Tycho
23\,M$_\odot$ is nearly as large as the difference between a Kepler
23\,M$_\odot$ and a Kepler 15\,M$_\odot$ star.  Unfortunately, the 
sharp boundaries in the Kepler models made it easy to predict 
the explosion energy or time (Fryer 1999).  Without these, estimating 
these explosion parameters becomes much more difficult. 

When our progenitor begins to collapse, we map the 1-dimensional star
into a 3-dimensional smooth particle hydrodynamics setup.  The
particles are added in a series of shells, where the number of
particles in each shell is determined by the density in the
1-dimensional progenitor and the mass of the particles.  The particle
masses are identical within each shell, but vary from shell to shell
in order to put the highest resolution in the inner portion of the
star near the action.  In each shell, the particles are placed in
random, but equally separated positions (see Fryer et al. 2006a for
details).  Although this randomness prevents any preferred direction
in the collapse, it does lead to density perturbations in the initial
model.  These density perturbations can be seen in an early time dump
of the collapsing star.  At high resolution, we can minimize these
pertubations, but for our low-resolution simulation, the magnitude of
the pertubation can become quite high.  We have lowered the tolerance
in the setup code to minimize this perturbation\footnote{The
shells are set up by putting particles randomly in a fixed shell and
then applying a repulsive force onto the particles until the deviation
in the separations falls within a given tolerance.  The shells are
then placed with random angles ($\theta,\phi$) with respect to one
another.  This allows a random distribution of particles.  By lowering
the tolerance, the shells are more evenly spaced.}.  Even so,
the initial density perturbation of this low-resolution calculation is
$\sim$3-5\%.  Note that this value is close to what we expect from
multi-dimensional models of the progenitors (Bazan \& Arnett 1998,
Meakin \& Arnett 2006a).  But our perturbations our not correlated,
and the multi-dimensional models predict more correlated
perturbations.  A correlated perturbation will likely produce an 
initial convective profile that is more asymmetric than our small-scale 
perturbations.  The larger asymmetries might lead to a stronger 
initial convection and more mixing in the ejecta of the explosion.

We chose a low resolution to delay the growth of convection.  It
has long been known that the SPH calculations by Herant et al. (1994)
and subsequent papers by Fryer and collaborators (e.g. Fryer \& Warren
2002) tend to develop strong convection earlier than models using grid
techniques.  Fryer \& Kusenko (2006) found that by limiting the
resolution, this delay can be mimicked in the SPH calculations.
Hence, the resolution is set to 500,000 particles for the entire
5.6\,M$_\odot$ stellar core modeled in our calculation.  We will
discuss the growth time of convection in more detail in the next
section (where we argue that the short delay in high-resolution 
SPH calculations is actually closer to reality than those calculations 
with delayed convection).

We use the SNSPH code (Fryer et al. 2006a) to follow the collapse,
convective, and ultimately explosion phase of this model.  This code
has passed several tests of its gravity routine.  Especially in
modeling the convective region when the neutron star begins to move,
it is important that this gravity routine be accurate.  The fact that
SNSPH is a gridless technique also makes it an ideal code for studying
neutron star motions, as no numerical issues arise when the neutron
begins to move.

SNSPH transport has been compared to 2-dimensional and 1-dimensional
flux-limited diffusion schemes.  But its neutrino transport is still
limited to a 3-flavor, single-energy flux-limited diffusion algorithm.
Such a scheme is believed to increase the net neutrino heating, making
an explosion easier.  Although the equation of state can couple an
accurate nuclear-statistical equilibrium algorithm (Hix \& Thielemann
1996) to the Lattimer-Swesty (1991) equation of state for neutron star
matter (see Herant et al. 1994 for details), to better compare to the
work of other authors, we use the Lattimer-Swesty equation of state
down to densities of $10^9 {\rm \, g \, cm^{-3}}$.  Because of the
incorrect energy levels in the Lattimer-Swesty algorithm for nuclear
statistical equilibrium, this choice alters significantly the entropy
profile of the convective region with our models (see Fryer \& Kusenko 2006)
\footnote{We note, however, that Janka et al. (2005) did not see any
change caused by a revised equation of state.  The different results
might be differences in the progenitor, differences in the algorithm
used for nuclear statistical equilibrium, or in the neutrino transport
algorihthm.  The importance of such revisions in the equation of 
state remains to be seen.}.

Smooth particle hydrodynamics is a Lagrangian code, and SNSPH uses the
standard artificial viscosity algorithm seen in many Lagrangian codes
to model shocks.  It tends to not model shock fronts as well as a grid
code using the piecewise parabolic method can do (at least as long as
the shock is traveling along the grid - see Fryer et al.  2006a for
details).  Also, this artificial viscosity is generally larger than
the true viscosity in core-collapse problems (at least those without
magnetic fields), effectively lowering the Reynolds number of our
numerical calculation\footnote{Algorithms exist that minimize this
damping effect (e.g. Balsara 1995).}.  This damps out high-order modes
in any convective instability.  We will discuss this effect in more
detail in the next section.

\section{Convective Instabilities}

50\,ms after bounce, an entropy gradient has developed just behind the
accretion shock (Figure 1).  Such an entropy gradient is extremely
susceptible to Rayleigh-Taylor convection.  One way to estimate the
timescale of this convection is to use the Brunt-V\"ais\"ala frequency
$\omega$ (see Cox, Vauclair, \& Zahn 1983):
\begin{equation}
\omega^2 = g/\rho (\partial \rho / \partial S)_{\rm P}
(\partial S/\partial r)
\end{equation}
where $\rho,S$ are the density, entropy of the matter, $(\partial \rho
/ \partial S)_{\rm P}$ is the partial derivative of the density with
respect to entropy at constant pressure, $(\partial S/\partial r)$ is
the partial derivative of the entropy with respect to the radius $r$
of that matter and $g \equiv G M_{\rm enclosed}/r^2$ is the
gravitational acceleration.  Here $G$ is the gravitational constant
and $M_{\rm enclosed}$ is the enclosed mass at radius $r$.  If
$(\partial S/\partial r)$ is negative, $\omega^2$ is negative and the
region is unstable.  The timescale for this convection ($\tau_{\rm
conv}$) is $(|1/\omega^2|)^{1/2}$.

In the limit where radiation pressure dominates the pressure term
(reasonably true at the accretion shock), this equation becomes:
\begin{equation}
\omega^2 = g/S (\partial S/\partial r) \approx (1/S) (G M_{\rm
enclosed})/r^2) (\Delta S / \Delta r)
\label{eq:timescale}
\end{equation}
where $\Delta S$ is the change in entropy over distance $\Delta r$.
Here we used the following relations: $S\propto T^3/\rho$ and ${\rm
Pressure} \propto T^4$.  For the conditions in Fig. 1, where $g
\approx 1.5 \times 10^{12} \, {\rm cm \, s^{-2}}$, $\Delta S/S \approx 0.2$,
and $\Delta r \approx 10^7 {\rm cm}$, the convective timescale is
roughly 2\,ms.  Even on core-collapse timescales, this is extremely
rapid.  It is worth noting that the negative entropy gradients in 
models using a modified equation of state (Herant et al. 1994, Fryer 
\& Kusenko 2006) have much larger amplitudes, leading to even more 
rapid growth of convection.

Some scientists prefer to estimate the growth time of Rayleigh-Taylor 
instabilities based on a more simplified equation using the Atwood number $A$:
\begin{equation}
\omega_{\rm Atwood}^2 = k g A
\end{equation}
where $A = (\rho_2-\rho_1)/(\rho_2+\rho_1)$ and $k$ is the wave number.  
Such an equation is designed for simplistic examples of a two density 
fluid chamber.  But, if we again assume a radiation pressure dominated 
gas, this equation becomes:
\begin{equation}
\omega_{\rm Atwood}^2 = k (G M_{\rm enclosed})/r^2) (\Delta S /S).
\end{equation}
If we pick a wave number roughly of the size scale of our convective 
region, this equation is identical to our equation derived using the 
Brunt-V\"ais\"ala frequency.

Figure 4 shows a time series with 6 snapshots of the convection in our
collapse core.  At 310\,ms, 10\,ms after the time in Figure 1, we see
that no vigorous convection has yet to develop.  50\,ms later, this
convection is strong and reaches beyond 200\,km.  The delay (beyond
the 2\,ms prediction of perturbation analysis) in this convection
could be because our rough estimate we obtained from
equation~\ref{eq:timescale} underestimated the timescale of
convection, but it is more likely that numerical viscosity from this
low-resolution calculation is damping the growth of the instabilities.
Fryer \& Kusenko (2006) found they could delay the convection by
pushing towards low resolution.  Fryer \& Kusenko (2006) artificially 
prevented an explosion by using low resolution to prevent the explosion.  
In this paper, we focus our study on late-time convection.  To ensure 
this, we use both a more massive progenitor, but also take advantage 
of the Fryer \& Kusenko (2006) result and use low resolution to delay 
the convection and, ultimately, the explosion.  We are intentionally 
damping the convection to allow us to study late-time convection.

Smooth particle hydrodynamics is not the only numerical hydrodynamics
technique that suffers from large numerical viscosity and it is likely
that the resolution (or lack thereof) in this calculation is causing
the delay in the convection.  Eulerian codes can also suffer from
numerical damping of convection through advection (Fryer et al. 2006c,
Schmidt et al. 2006).  

\subsection{Understanding Low-Mode Convection}

Based on the low-mode convection from the Herant et al. (1994)
2-dimensional simulations, Herant (1995) argued that in the extreme
case where the modes ultimately merged to produce convection along a
single mode could produce asymmetric explosions and neutron star
kicks.  However, subsequent SPH calculations never developed single
mode convection (Fryer 1999, Fryer \& Warren 2002, Fryer \& Warren
2004, Fryer 2004, Fryer \& Kusenko 2006) unless some initial asymmetry
(neutrino-driven kick, asymmetry in the progenitor, rotation) was
placed on the collapse to seed this $l=1$ mode.  However, recently, a
number of simulations exhibit strong $l=1$ structures (Blondin et
al. 2004,2006; Scheck et al. 2004; Burrows et al. 2006).  Although
these simulations argue for this low-mode convection, we have yet to
put together a complete picture of the physics affecting this
convective region.  As such, it remains difficult to extract the
numerical effects from those effects of a true physical nature.  There
exist a number of similarities between the convective instabilities in
the core-collapse problem with those of Bondi-Hoyle-Littleton
accretion (see Foglizzo et al. 2005 and references therein) and, 
already, this knowledge is being applied to the core-collapse problem.

What we do know is that there are many possible drivers behind this
convection.  First and foremost, as we showed above, the entropy
profile in the convective region is extremely unstable to
Rayleigh-Taylor instabilities.  If the accretion shock capping this
region moves outward, the entropy gradient will continue to be
produced simply because the entropy jump at the shock decreases as the
shock moves outward (Houck \& Chevalier 1991, Fryer et al. 1996).
Neutrino heating (and shock heating as the downflows strike the
proto-neutron star) from below also serves to maintain the entropy
gradient and drive Rayleigh-Taylor convection.  Herant et al. (1994)
focused their discussion on this convective instability.  One would
expect such convection to produce bubbles with sizescales roughly on
the radial extent of the convective region ($\sim r_{\rm outer
accretion shock} - r_{\rm proto-neutron star}$).  When the region is
small, we expect many downflows and upflows.  As the convective region
pushes the accretion shock outward, the number of downflows should
decrease.  It is this trend that we see in our simulations
(Fig.~\ref{fig:conv1}).

A second instability has been studied by Blondin et al. (2004,2006),
focusing on the accretion instability caused by spherical accretion.
This instability tends to drive low order $l=0,l=1$ modes and Blondin
et al. (2006) found that such modes do dominate in conditions where
Rayleigh-Taylor convection is not strong.  It is interesting to note
that this spherical accretion shock instability was discussed in
detail by Houck \& Chevalier (1992) to study accretion onto neutron
stars.  Fryer, Herant, \& Benz (1996) modeled this accretion in
2-dimensions and found that the Rayleigh-Taylor instabilities set up
by the entropy gradient produced as the accretion shock moved outward
again dominated the instabilities.  They found that $l=$few, not $l=0$
or $l=1$, modes dominated the convection at early times.  $l=$ few
modes is what one expects from Rayleigh-Taylor convection and Fryer et
al.  (1996) believed this dominated the convection\footnote{But bear
in mind that the Fryer et al. (1996) calculation was limited to a
2-dimensional calculation in s 90$^\circ$ wedge, so we should take
these calculations with a grain of salt.}.  In our collapse
simulation, we might also expect that Rayleigh-Taylor convective
instabilities to dominate the matter motion at early times.  But as
the convection persists, we see the development of an $l=1$ mode that
is probably caused by the accretion shock instability studied by
Blondin \& Mezzacappa (2006).  It appears that a combination of these
two instabilities can explain the matter motion in our simulation.

A third instability, originally highlighted by Blondin et al. (2004)
has also piqued the curiosity of the core-collapse community: the
vortical-acoustic instability.  This instability, which takes the
vorticity pulled down in the downflows and converts it to sound waves,
drives low-mode convection.  Although Blondin \& Mezzacappa (2006) now
believe the instabilities they see are not the vortical-acoustic
instability, new adherents (e.g. Burrows et al. 2006) have brought
continued support to this particular instability.  Foglizzo et
al. (2006), Ohnishi et al. (2006) and Yamasaki \& Yamada (2006) have
also studied the relative importance of this instability.  Ohnishi et
al. (2006), in particular, pointed out that the relative importance of
these shock instabilities will depend upon exact conditions in the
models and might vary for different progenitors.  The convection in
our calculations can either be explained by Rayleigh-Taylor plus this
vortical-acoustic instability, or by Rayleigh-Taylor plus late-time
accretion shock instabilities argued by Blondin \& Mezzacappa (2006).

In addition to differences in progenitors, which of these instabilities 
dominate in actual calculations may well be a reflection of
numerical technique and not what nature produces.  As an example of
the role of numerics, let's analyze the laminar nature of our
downflows.  In our calculations, we set the the SPH viscosity
parameters $\alpha,\beta$ to 1.0,2.0 respectively.  With our low
resolution calculation and the velocities of and sound speeds in the
downflows, we find that our numerical Reynolds number is $\sim 15$.
Fryer \& Warren (2004) did produce one simulation where the Reynolds
number was closer to 100, but no SPH calculation of core-collapse
supernovae has modeled significantly higher Reynolds numbers.  As
such, we expect all of our simulation to exhibit laminar flows, and
they do.  If convection is important in the supernova explosion, 
we must make sure that our numerical models match the conditions in nature.

What do we expect from nature?  If the viscosity is dominated by the
Spitzer viscosity (Braginskii 1958, Spitzer 1962), then the Reynolds
number is many orders of magnitude higher than what we model.  If this
were the only viscosity, then nature would produce turbulent flows and
the entire structure of the convection would be different.  Currently,
most simulations exhibit laminar (or close to laminar flows).  Either
these simulations, like our SPH simulations, have enough numerical
viscosity to effectively be modeling laminar flows, or some other real
viscosity is playing a role in reducing the Reynolds number.  It is
possible that magnetic fields can add viscosity and effectively reduce
the Reynolds number, but most collapse simulations to date do not
include magnetic fields (Simon 1949, Spitzer 1962).

Colgate (pvt. communciation 1996) suggested that neutrinos could
increase the viscosity.  Recall that the zeroth order Spitzer 
viscosity is given by:
\begin{equation}
\eta^e_0 \propto n k_{\rm Boltz} T_e \tau_e
\end{equation}
where $n$ is the number density, $k_{\rm Boltz}$ is the Boltzmann
constant, $T_e$ is the electron temperature and $\tau_e$ is the
collision time of the electrons:
\begin{equation}
\tau_e \propto T_e^{3/2}/(n \lambda) \,{\rm s}
\end{equation}
where $\lambda$ is the Coulomb logarithm.  This gives us the familiar
$T^{5/2}$ dependence of the Spitzer viscosity.  Let us understand this
viscosity term a little bit better.  The viscosity is essentially
determined by the ability for the electron/ion/particle to transport
momentum.  This depends upon 2 factors: (1) how much momentum the
particle contains (the specific energy is the specific momentum
squared) and (2) how far the particle transports this momentum.  If
there is not much energy in the particles, they do not contribute much
to the viscosity.  If the particle does not move significantly before
losing its preferred direction, it also does not contribute much to
the viscosity.  This latter effect is what reduces the contribution
from electrons to the viscosity and this is why we expect a large
Reynolds number if we constrain ourselves only to the the electron
viscosity.  Colgate's idea was that the neutrinos, while trapped at
the base of the convection, have a much longer collision time than the
electrons.  Using the above equations to estimate the viscosity from
neutrinos (with roughly 1-10\% the energy stored in electrons, but 
with a collision time that can be 10 orders of magnitude longer than 
electrons), we find that the neutrino viscosity can reduce the Reynolds
number in our simulated downflows down to 1000 (and perhaps 100).  For
such cases, laminar-like flows are expected.  But such features must
be checked in every simulation.

Numerical versus real viscosity is just one example of how we have to
be careful to distinguish between the results of our simulations and
what is actually happening in nature.  As convection plays a larger
role in understanding supernovae, real versus numerical becomes a very
important question that must be addressed.  In the case of our current
simulation, the laminar flows may well be real and the nature and
evolution of the convection can be understood by our understanding of
Rayleigh-Taylor and accetion shock instabilities.  We do know that
the dominant modes of the convection are set by the driving forces,
which is not too dependent on the Reynolds number.  The coherence of
the flows would be changed by the onset of turbulence, but this might
be a minor effect on the supernova engine.  But we are also certainly
seeing numerical delays on the onset of convection and the numerical
viscosity in the code is setting the Reynolds number of the
calculation.  Fryer \& Kusenko (2005) claim that these numerical 
delays are very important and comparing their results with 
with the Fryer \& Warren 2002 results, this statement, as far as 
smooth particle hydrodynamics simulations are concerned, is true.   
Whether or not such numerical artifacts are affecting Eulerian 
calculations awaits convergence study calculations done by these 
groups.

\section{Proto-Neutron Star Motion}

Figure~\ref{fig:ns1} shows 4 snapshots in time of the central 200\,km
of our calculation.  The proto-neutron star has moved slightly before
the onset of convection due to slight asymmetries in the collapse
conditions.  But the real motion occurs after convection becomes
strong and downflows ``kick'' the neutron star.  The net kick arises
as convective downflows flow down and strike the proto-neutron star,
giving it a series of ``mini-kicks''.  Calculations using the same
code used here, but with a 15 instead of 23\,M$_\odot$ star, higher
resolution, and the Herant et al. (1994) coupled equation of state did
not exhibit this motion without using large seeds in the collapsing
core (see discussion in Fryer 2004).  This is almost
certainly caused by the delay in the explosion, as was suggested by
Scheck et al. (2004).  Note that the gravity solver in SNSPH is
ideally suited for motion of the neutron star and has been shown to
behave well in such difficult gravitational calculations (Fryer et
al. 2006a), so we believe these motions are real (with the caveat that
the small perturbations in the initial model, although representative
of what we expect of collapse progenitors, is artificially placed in
our initial conditions).

The x,y,z position of the proto-neutron star center-of-mass is shown
in Fig.~\ref{fig:nspos}.  Here we have defined the proto-neutron star
as all matter with densities in excess of $10^{13}\,{\rm \, g \,
cm^{-3}}$.  The motion of the neutron star is not monotonic, but there
is a basic trend in its motion.  In Fig.~\ref{fig:nsvel}, the velocity
components of this neutron star clearly show why such oscillations in
the positions occur.  The proto-neutron star velocities oscillate with
amplitudes above 100\,km\,s$^{-1}$, but these large amplitudes occur
over 100\,ms time periods.  The power in these oscillations is less
than $5\times10^{48} {\rm erg \, s^{-1}}$, much lower than that
predicted by Burrows et al. (2006).  The velocity oscillations 
are caused by the time evolution of the downflows and the location 
at which they strike the proto-neutron star.

A pressure wave could move through the star without significantly
moving any of the matter.  To test if the pressure in a shell of
matter is varying widely with time, we plot the time evolution of 9
particles 35\,km from the center of the neutron star
(Fig~\ref{fig:osci}).  The time resolution in this plot is roughly
0.04\,ms.  At the 0.04\,ms timescale, we do not see large oscillations
in the pressure of the particles.  The particles slowly compress, and
there is some variation at timescales comparable to our downflow
timescales, but no obvious ringing.  We must now try to understand why
we do not observe the ringing of the neutron star found in the Burrows
et al. (2006) models.  It is possible that the smooth particle
hydrodynamics technique is damping out any possible pressure
perturbations.  It is also possible that real damping (e.g. Silk
damping) is preventing these oscillations (Ramirez-Ruiz, private 
communication).

Figure~\ref{fig:nsmass} shows the mass of the compact remnant as a
function of time.  The compact remnant steadily accretes mass during
the convective phase as material piles up and cools onto the neutron
star.  By the time the explosion occurs, the baryonic mass of the
remnant exceeds 1.8\,M$_\odot$.  Because the explosion is weak ($<
10^{51}$\,erg), this remnant will accrete further through fallback
and will certainly form a black hole.  The large mass is a common
feature in delayed explosions for progenitors stars more massive the
15\,M$_\odot$.  Stars above this mass that have such long delays in
their explosion will not produce generic neutron stars.  Recall 
that the long delay is, in part, caused by low resolution in the 
convection and, in part, by the massive progenitor.

The downflows also carry a modest amount of angular momentum down into
the neutron star.  The top panel in figure~\ref{fig:nsspin} shows the
absolute value of the x,y, and z components of the specific angular
momentum in the proto-neutron star as a function of time.  The bottom
panel shows the x,y, and z components of the angular velocity.  Note
that the direction of the spin changes as a function of time, but 
the spin never reaches periods below 15\,ms and 1\,s spin periods are 
more typical.  But it is possible that spin periods in the tens of 
milliseconds can be achieved even if the progenitor itself is not rotating.

During the convection, the neutron star is, at times, moving rapidly
and, at times, spinning rapidly.  But how are these two related.
Figure~\ref{fig:velspin} shows the relationship between the kick
velocity and the neutron star spin.  The top panel shows the angle
between the velocity and spin vectors ($\mathbf{v} \cdot
\mathbf{\omega}/|\mathbf{v}|/|\mathbf{\omega}|$) as a function of
neutron star velocity at all points in time.  The bottom panel shows
the magnitude of the spin versus the neutron star velocity.  Depending
upon when the explosion occurs, we can obtain a range of results.  One
might expect, since it is the accretion of the downflows that both
produces the kick and the spin, the two values might evolve together.
However, with this single model, it does not appear that there is any
correlation between velocity and spin, nor of the direction of the
spin with respect to the velocity.  More simulations are required to
determine whether this lack of correlation is a property of this 
explosion mechanism.

\section{Implications}

\subsection{On Supernovae}  
In the collapse of a 23\,M$_\odot$ star with our SNSPH code using low
resolution, we obtain an extremely weak explosion at late times
(400\,ms after bounce).  In this delayed explosion, we see effects
from both Rayleigh Taylor and spherical accretion shock instabilities.
We do not produce a dominant $l=1$ mode in our convection, but it is
present at late times just prior to the explosion.  The convective
downflows ``kick'' the proto-neutron star, giving it velocities as
high as 150\,km\,s$^{-1}$.  This is comparable to the low velocity set
of simulations by Scheck et al (2004).  Although our neutron star does
not receive a large ($>450$\,km\,s$^{-1}$) kick, we can not rule out
such large kicks with mildly different initial conditions.  Indeed, it
is likely that different conditions (both in the initial star and the
numerical setup) could produce larger kicks (e.g. Fryer 2004; Scheck
et al. 2004).

Our proto-neutron star is definitely buffeted by the downflows.  And
although these downflows do impart a kick onto the neutron star, we
do not see any ringing or oscillations in the proto-neutron star.
This result does not agree with the recent work of Burrows et
al. (2006).  A number of reasons could cause this difference.
Numerical viscosity in the particles making up the proto-neutron star
could be damping out these oscillations.  But the difference may be
due to the fact that our simulation did not exhibit the dominant $l=1$
mode convective cycle seen by Burrows et al. (2006).  It may be that
this dominant mode is necessary to drive oscillations.  Fortunately,
such questions can be studied using perturbation analysis coupled with
a detailed understanding of the equation of state (Arras et al.
2006).  In any event, both our convective engine and that of Burrows
et al. (2006) takes far too long for this progenitor to make a normal
neutron star.  It is likely that both explosion mechanisms will produce 
black holes.

Because the explosion takes so long to occur, the neutron star cools
with a heavy mantle of material on top of it.  This mantle ultimately
places all the neutrinosphere (radius of last scattering) for the
electron neutrino and electron anti-neutrinos at roughly the same
position.  Hence, the energies of these two neutrino species are
nearly identical.  Figure~\ref{fig:neut} shows the evolution of the
neutrino energies and luminosities for the 3 species followed in this
calculation: electron neutrino ($\nu_e$), electron anti-neutrino
($\bar{\nu}_e$), and all others ($\nu_x \equiv$ $\mu,\tau$ neutrinos
and anti-neutrinos).  The electron neutrino energy is under 12\,MeV
whereas the electron anti-neutrino energy is under 14\,MeV.  Such
small differences between the neutrino energies is consistent with
many of the other delayed supernova explosions (e.g. Buras et
al. 2006).  Note that the flux-limited diffusion calculation tends to
overestimate the mean neutrino energy by $\sim$10\% (Budge et
al. 2006).  The $\sim$ factor of 2 higher flux coming out in
electron neutrinos (with energies that are only $<$20\% lower) means
that matter above the neutrinosphere will preferentially absorb
electron neutrinos, leading to an increase in its electron fraction.
This will play a role in the nucleosynthetic yields (studied in a
later paper) from this explosion.

Neutron star spins at the tens of milliseconds level are possible 
even with a non-rotating progenitor.  For spin-periods below 10\,ms, 
it is likely that a rotating progenitor is required.  

The entropy profile of our star 600\,ms after collapse is shown in
figure~\ref{fig:entr}.  Even at these late times, the peak entropies
do not rise above 20\,$k_{\rm B}$ per nucleon.  Such low entropies
would not be high enough to produce the r-process in a wind-driven
trajectory.  But the trajectories in this convective region are very
different from the wind-driven trajectories, and the trajectory can be
more important in determining the actual yield of a piece of matter
than the entropy or electron fraction in that matter (see, for
example, Meyer 2002 or Fryer et al. 2006d).  We defer discussion of the 
exact yields of this collapse to a later paper.

It has been over a decade since the first collapse and explosion
calculations in 2-dimensions suggested that convection play a crucial
role in the core-collapse supernova engine.  We now know several
different convective instabilities (even without including the effects
of magnetic fields).  To truly understand the role these instabilities
play in supernovae, we will have to understand the limitations of our
numerical techniques.  Without much higher resolution, or new
techniques for including artificial viscosity, our SNSPH code is
limited to low Reynolds number flows.  Most grid codes suffer from
similar viscosities and peculiarities due to advection terms.
Techniques for solving gravity and estimates for the equation of state
also can alter the convection.  If convection is an important
ingredient of the supernova engine, we have our work cut out for us.

\subsection{On Hypernovae and Black Holes} 

Fryer (1999) argued that 23\,M$_\odot$ stars lie at the transition
between neutron star and black hole formation.  As such, these stars
are subject to a range of outcomes.  The delay in the explosion allows
time for large magnetic fields to develop and it is possible that if
this star were rotating, large magnetic fields would develop that
could dominate the explosion.  In addition, as the proto-neutron
star's mass increases, it is possible that a transition to quark
matter can occur, producing an explosion.  Lastly, if the star
collapses to a black hole, it can produce an explosion via a black
hole accretion disk engine (Popham et al. 1999).  The fact that 
we are in this transition region means that a wide range of 
explosive fates exist for these stars.

Without magnetic fields or a transition to a quark star injecting
energy, this star will ultimately accrete so much material that it
will collapse to form a black hole.  If the star were rotating, it
would be a candidate star for the production of a hypernova (and even
the subclass of hypernovae that produce gamma-ray bursts).  These
calculations have some important implications for the hypernova
engine.  In this simulation, the collapse initially produces a massive
neutron star that, through fallback, collapses to form a black hole.
In such a scenario, we will get a weak supernova explosion followed
within a few seconds of a collapse to a black hole and a hypernova jet
explosion.  Depending upon this delay, the $^{56}$Ni produced can vary
dramatically (Fryer et al. 2006b).  This delay also allows the massive
newly-formed neutron star to move away from the ``center'' of the
star.  Recall that we found kicks as high as $150 \,{\rm km \,
s^{-1}}$ in our calculation (and the Sheck et al. (2004) results very
large kicks $1000 \,{\rm km \, s^{-1}}$ are possible).  If the
hypernova engine does not turn on until a few seconds after collapse,
the black hole may well be over 1000\,km away from the star's center
when it turns on.

Even if the explosion does not occur and the the core collapses
directly to a black hole, we must still pass through a phase of
long-term convection.  This convection will very likely develop low
mode convection (this depends upon how far the convective region moves
outward) and the compact remnant will be kicked.  In the most
conservative case where there is no matter carrying out momentum,
asymmetric neutrino emission will carry out momentum to produce a kick
on the black hole, and it is likely that all (even direct-collapse)
black holes are born with moderate kicks.  Past estimates (e.g. Fryer
) have suggested that black holes receive kicks with comparable
momenta to the neutron star velocity distribution ($v_{\rm BH}/v_{\rm
NS} = M_{\rm NS}/M_{\rm BH}$).  What we expect from these models (and
assuming this asymmetric convection is the source of kicks on compact
remnants) is instead that the black hole velocity distribution is
comparable to the neutron star distribution.  The affect of this kick
distribution on black hole binary systems remains to be seen.

\acknowledgements {\bf Acknowledgments} We would like to thank many
useful conservations with Bill Rider, Rob Lowrie, Gary Diltz, Enrico
Ramirez-Ruiz, Phil Arras, Dave Arnett, Paolo Mazzali, Philipp
Podsiadlowski, Brian Schmidt and Ken Nomoto.  This work was funded in
part under the auspices of the U.S.\ Dept.\ of Energy, and supported
by its contract W-7405-ENG-36 to Los Alamos National Laboratory, by a
NASA grant SWIF03-0047, and by National Science Foundation under Grant
No. PHY99-07949.

{}

\clearpage

\begin{figure}
\plotone{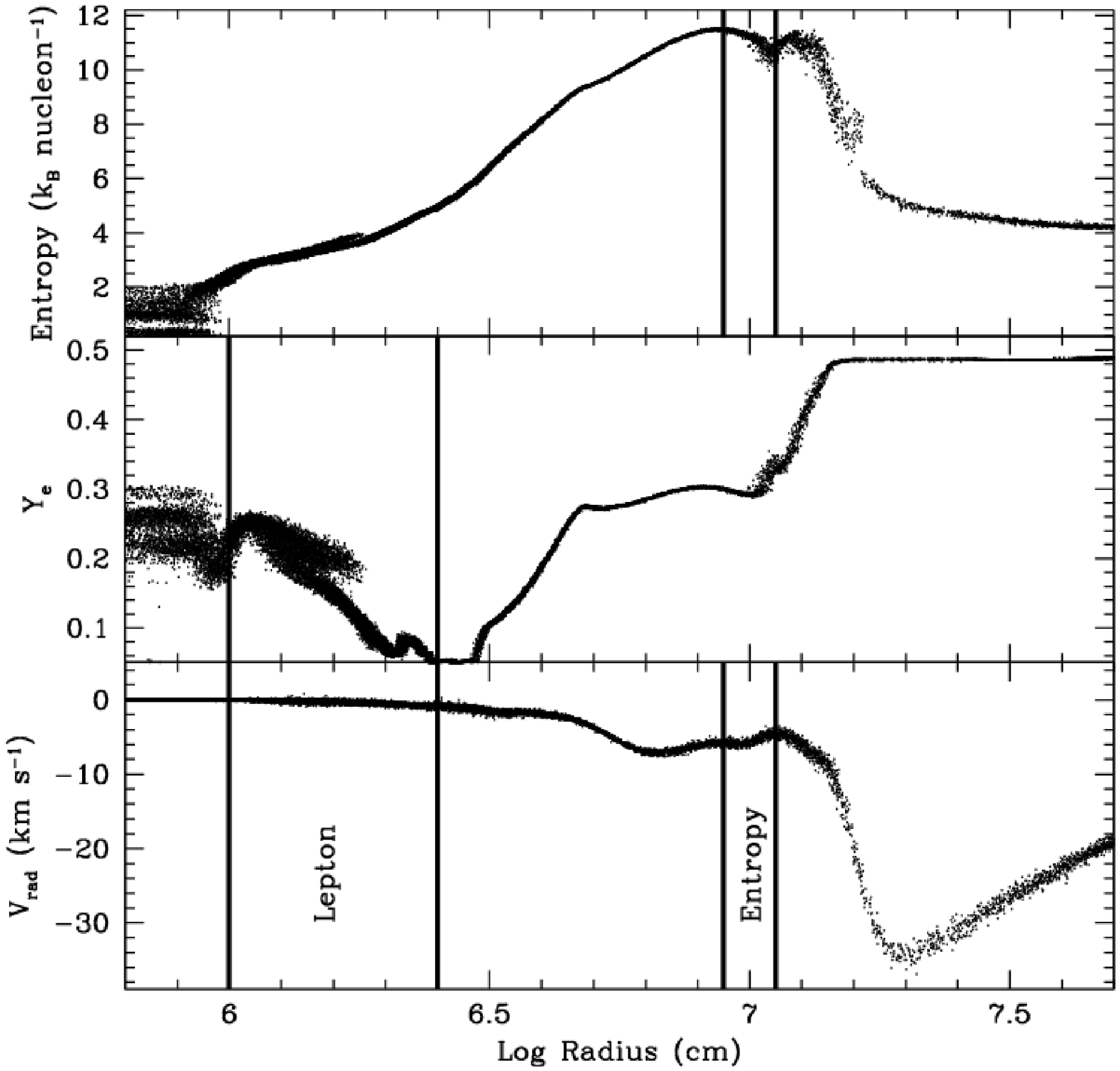}
\caption{Entropy (top), Electron Fraction (Middle), and Radial
Velocity (bottom panel) as a function of radius for our simulation
50\,ms after bounce (roughly at the time of the stall of the bounce
shock).  We see that a negative lepton gradient has developed near the
neutrinosphere.  It is this initial gradient that leads to convection
within the proto-neutron star.  Beyond the neutron star, just below
the accretion shock, we see the initial negative entropy gradient that
initially drives the convection in this outer region.  At this time,
with this low resolution simulation, no strong convection has
developed, but these are the two regions in which we expect 
convection to occur.}
\label{fig:diag}
\end{figure}
\clearpage

\begin{figure}
\plotone{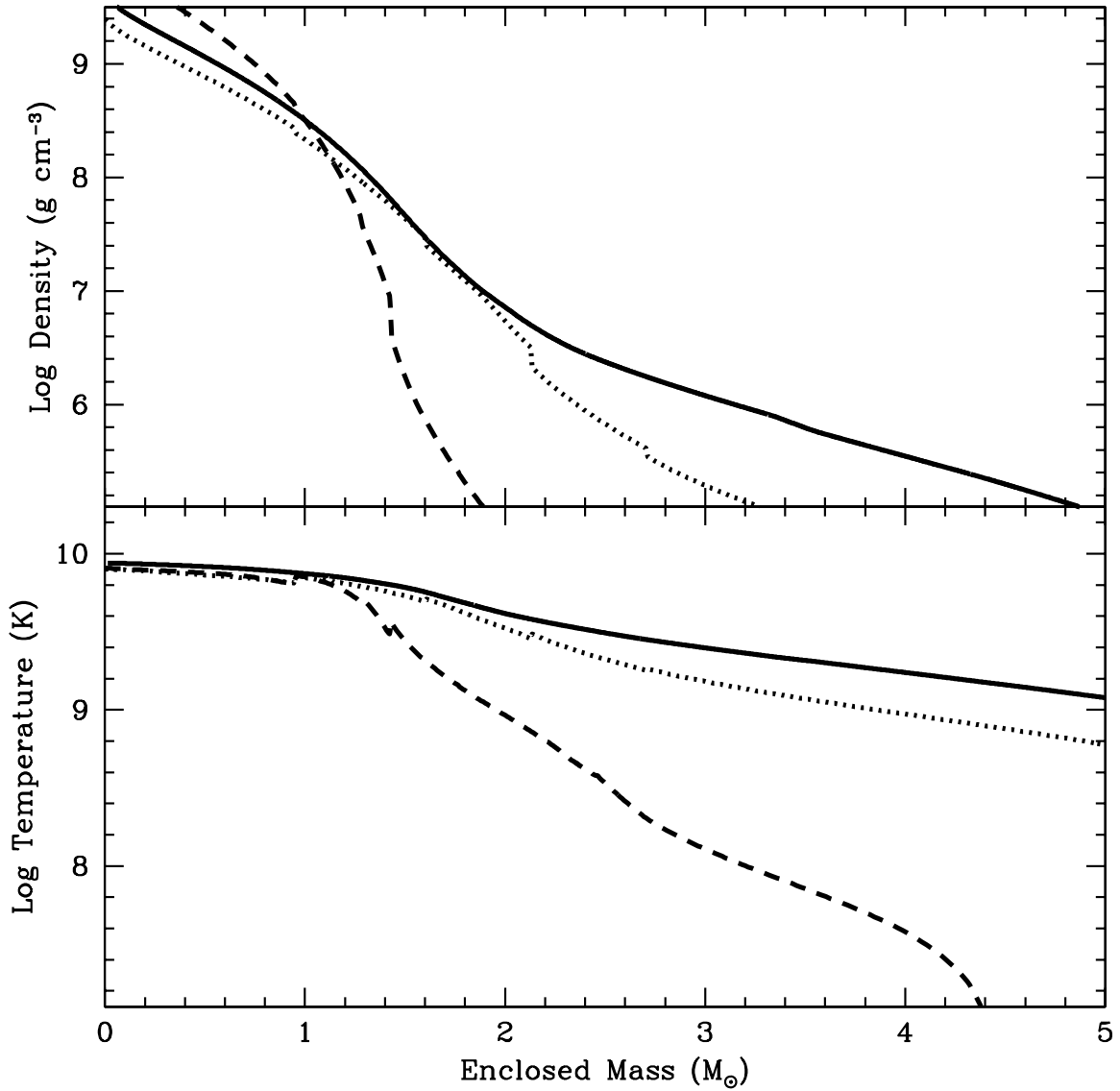}
\caption{Density (top) and Temperature (bottom) as a function 
of enclosed mass for our progenitor (solid line:  Young et al. 2006), 
a KEPLER progenitor (dotted line:  Heger et al. 2006), and the standard 
s15s7b2 model (dashed line:  Woosley \& Weaver 1995).  Note the jumps 
in density just above 2.1\,M$_\odot$ and 2.7\,M$_\odot$.  This occurs 
because of the sharp transition between shells in the KEPLER code.  It 
does not occur in our progenitor.  Our progenitor also has higher 
densities beyond about 1.5\,M$_\odot$.}
\label{fig:stellar1}
\end{figure}
\clearpage

\begin{figure}
\plotone{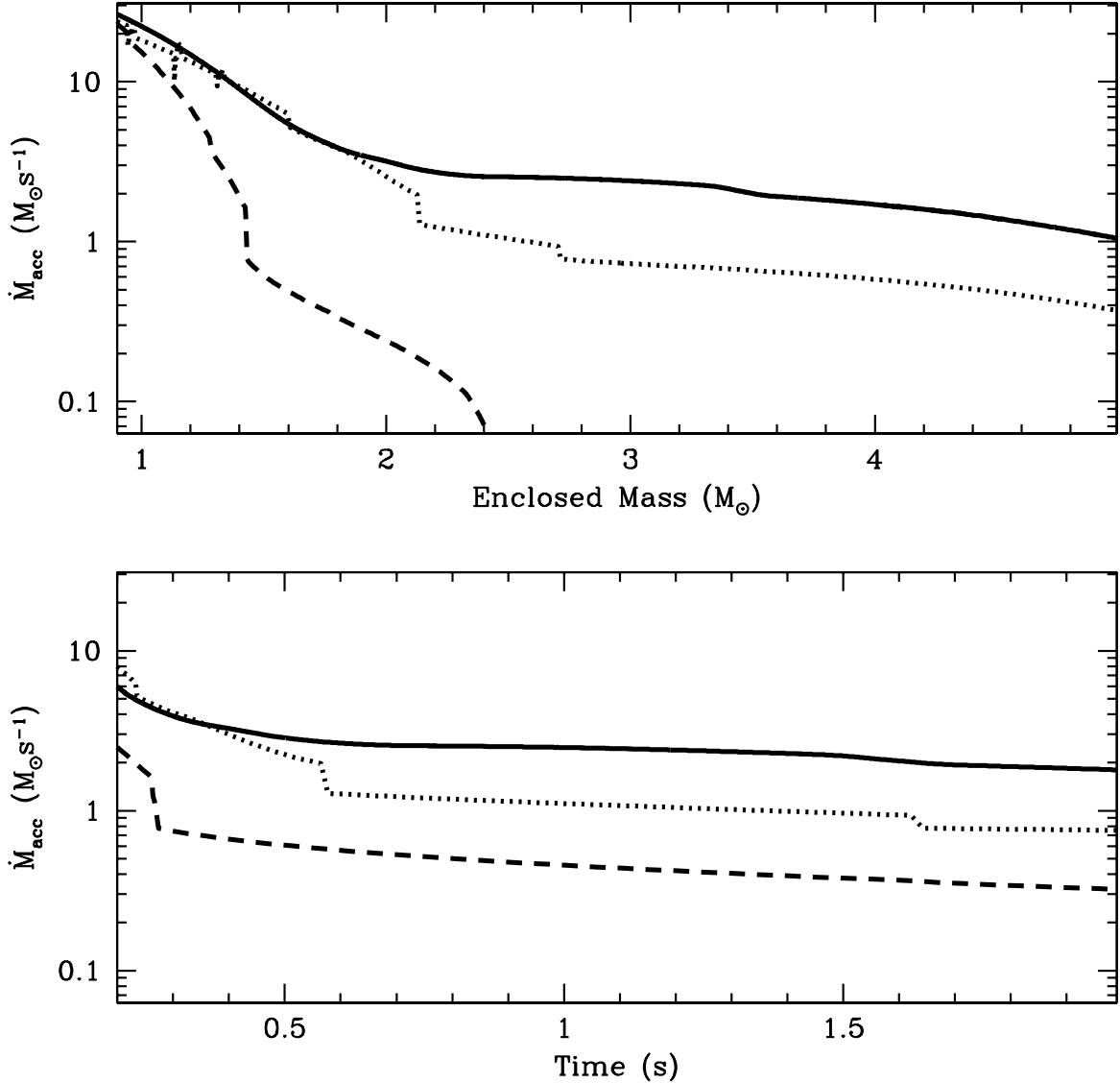}
\caption{Accretion rate as a function of enclosed mass (top) and time
(bottom) for 3 progenitors: our progenitor (solid line: Young et
al. 2006), a KEPLER progenitor (dotted line: Heger et al. 2006), and
the standard s15s7b2 model (dashed line: Woosley \& Weaver 1995).
This accretion rate is a good simple (but not complete) indicator of
the fate of the star where stars with higher mass accretion rates are
harder to explode.  Note that at 0.65\,s (roughly 0.4\,s after
bounce), the difference between the KEPLER model and our progenitor
(both 23\,M$_\odot$ stars) is roughly equal to the difference between 
the KEPLER model and model s15s7b2.  Clearly, the uncertainty in stellar 
progenitor codes is still large.}
\label{fig:stellar2}
\end{figure}
\clearpage

\begin{figure}
\plotone{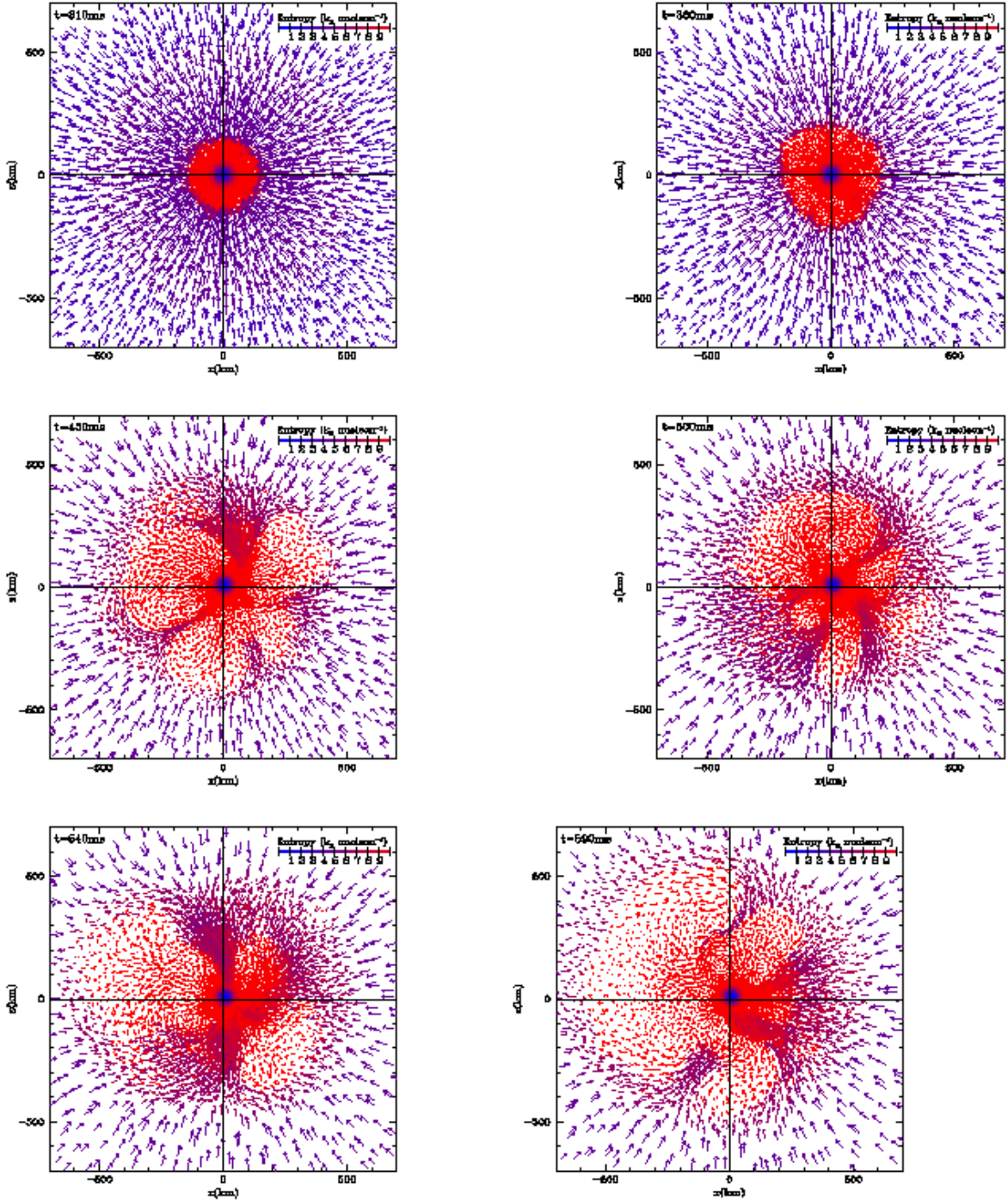}
\caption{6 snapshots in time of the convection in 
our collapsing model.  We plot slices of the data 
in the x-z plane.  The vectors denote direction 
and magnitude of the particle motion.  The colors 
denote entropy.  Probably because of our low resolution, 
the growth time is longer than we would expect from 
an instability analysis.  But convection does develop, 
ultimately producing a weak explosion.  The convection 
also is far from symmetric, but we do not get the 
single-sided downflows seen in many recent 2-dimensional 
calculations.}
\label{fig:conv1}
\end{figure}
\clearpage

\begin{figure}
\plotone{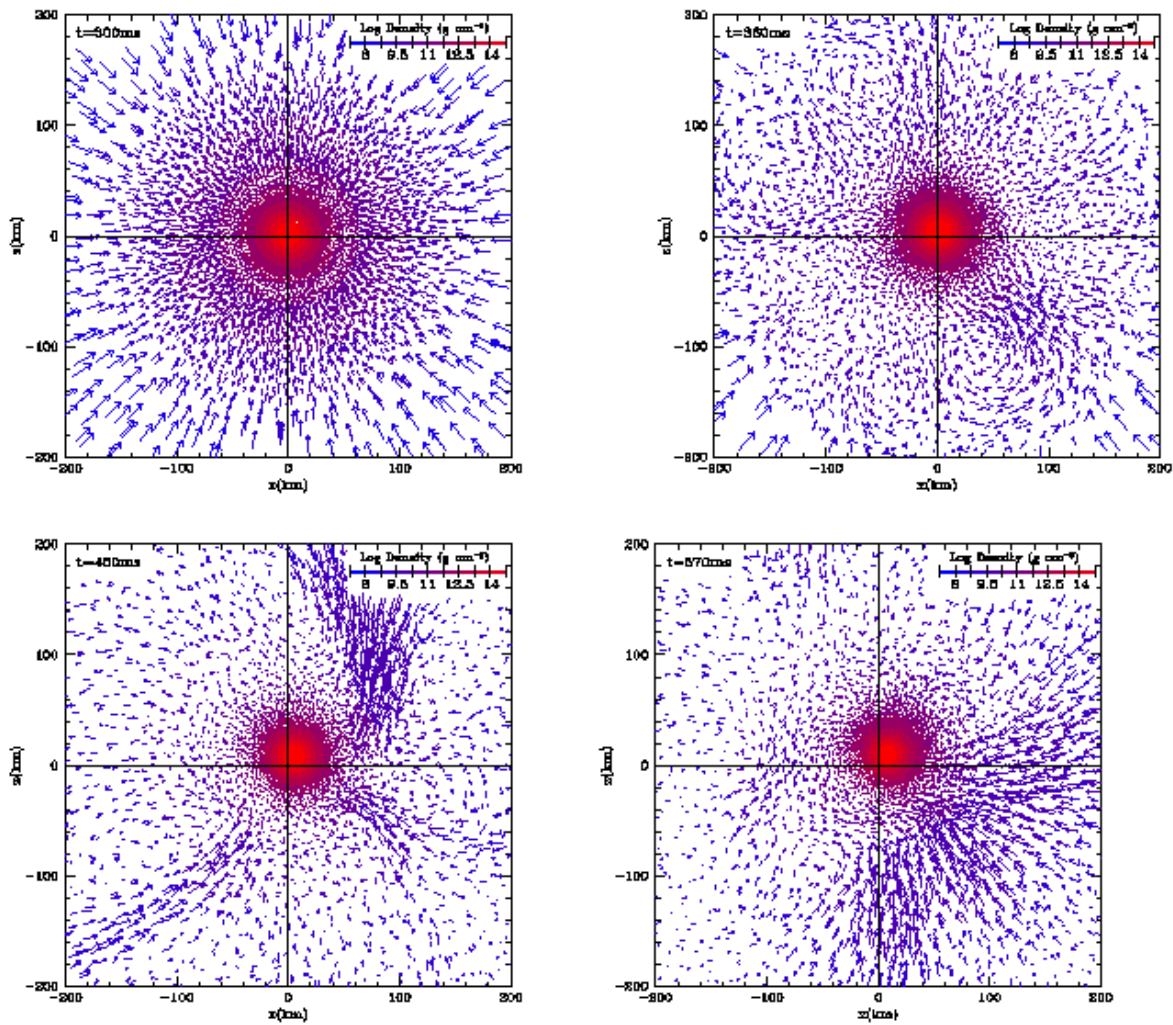}
\caption{4 snapshots in time of the central region surrounding the
neutron star.  We plot slices of the data in the x-z plane.  The
vectors denote direction and magnitude of the particle motion.  The
proto-neutron star does not move considerably until convection
produces downflows that impart ``mini-kicks'' onto the proto-neutron
star.}  
\label{fig:ns1}
\end{figure}
\clearpage

\begin{figure}
\plotone{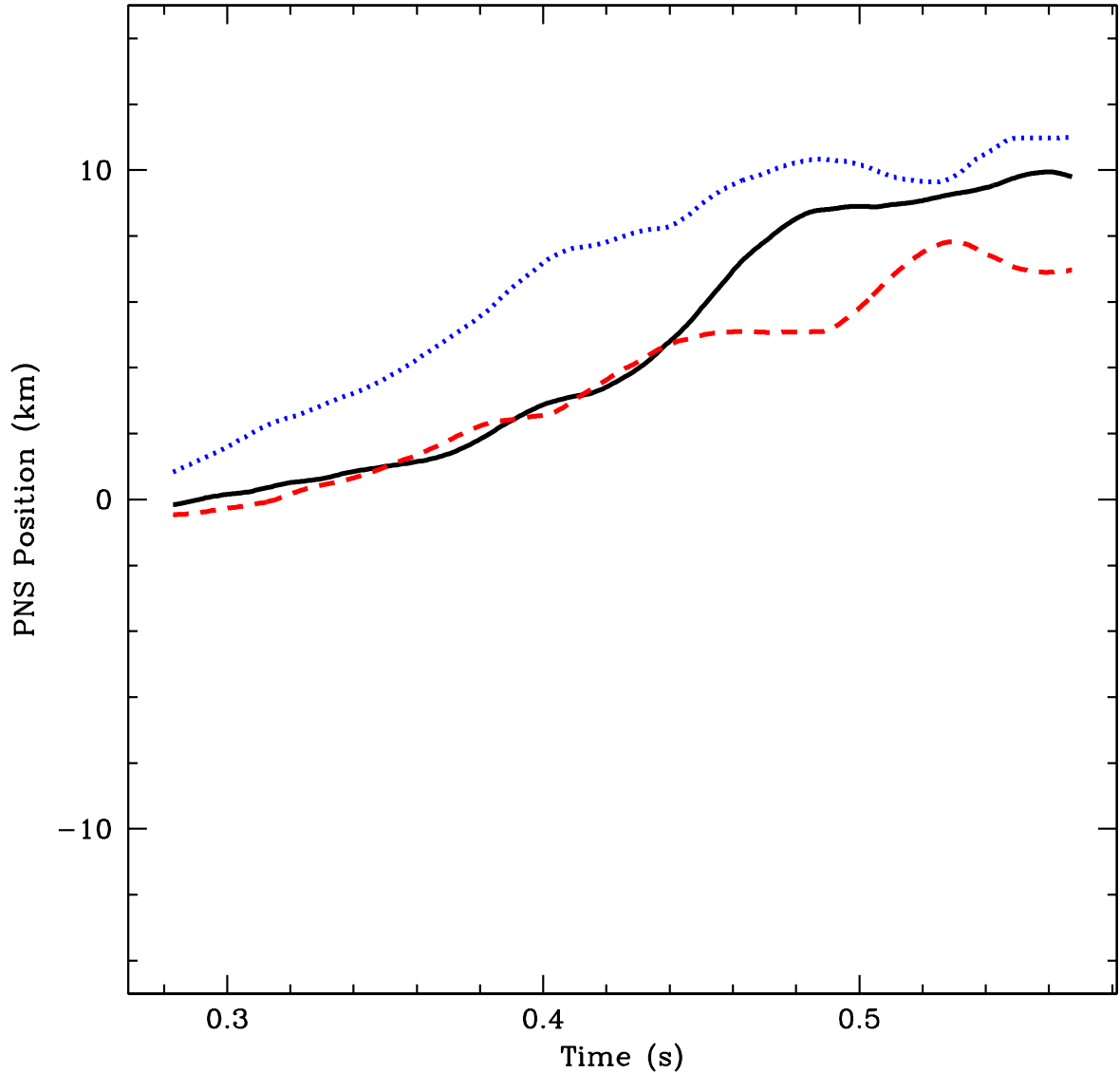}
\caption{The x (solid), y (dotted), and z (dashed) center of 
mass positions of the proto-neutron star as a function of time.  
We have defined material in the proto-neutron star as that material 
whose density exceeds $10^{13} {\rm g cm^{-3}}$.  The motion is 
not monotonic, but driven by the downflows that buffeting the 
proto-neutron star.  However, there is a systematic trend in the 
motion of the neutron star, initially seeded by a slight density 
perturbation in the collapsing star.} 
\label{fig:nspos}
\end{figure}
\clearpage

\begin{figure}
\plotone{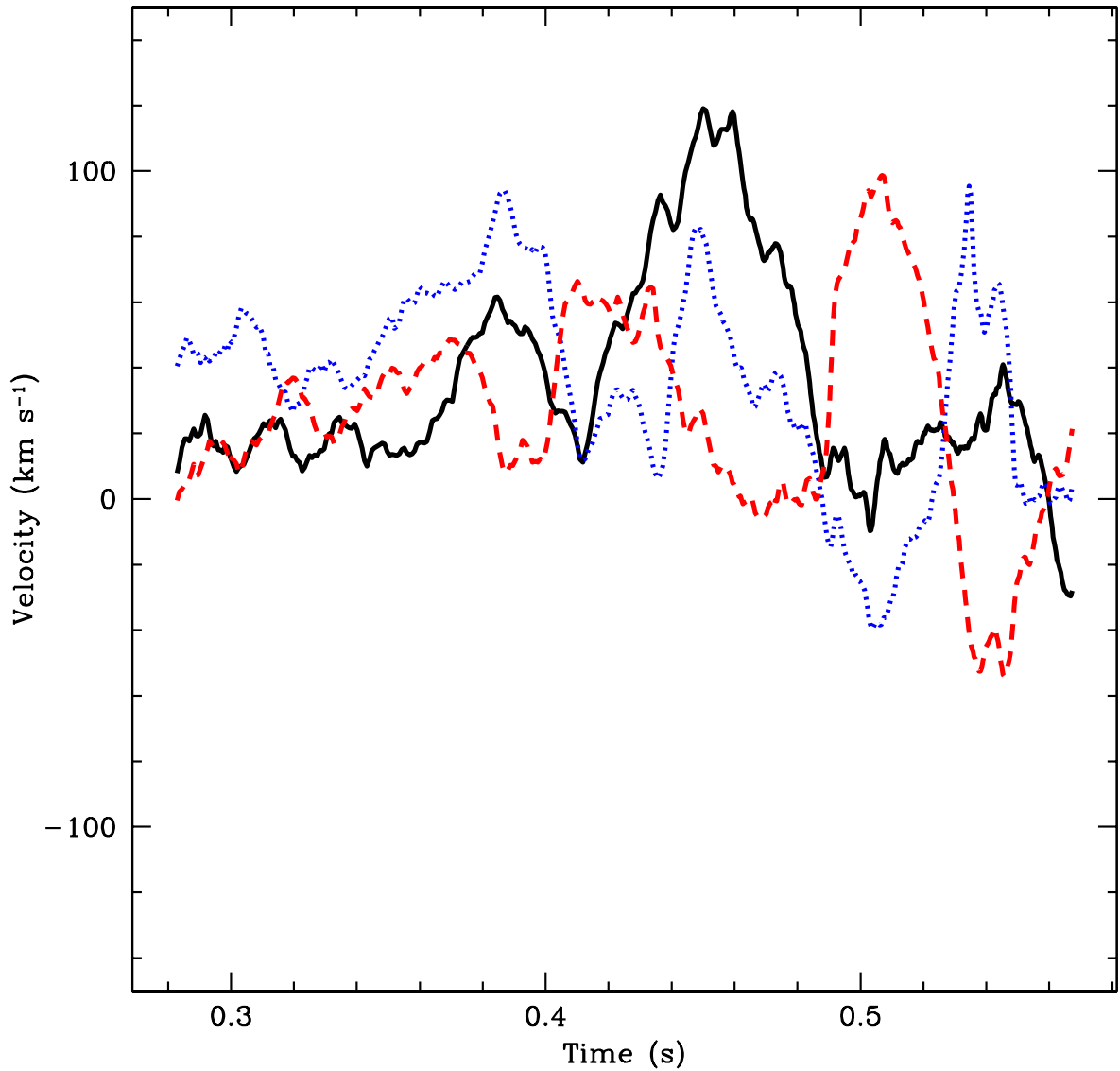}
\caption{The x (solid), y (dotted), and z (dashed) velocities of the
proto-neutron star as a function of time.  We have defined material in
the proto-neutron star as that material whose density exceeds $10^{13}
{\rm g cm^{-3}}$.  The velocities are definitely erratic caused by 
the kicks imparted by downflows onto the proto-neutron star.  Because 
no low-order mode develops and the downflows do not impart cumulative 
kicks.  In this calculation the proto-neutron star velocity never 
exceeds $\sim 150 {\rm km s^{-1}}$, but the velocity is comparable 
to the low-velocity subset of the Scheck et al. (2004) calculations.  
Note that we don't see any rapid oscillations of a ringing neutron star.}
\label{fig:nsvel}
\end{figure}
\clearpage

\begin{figure}
\plotone{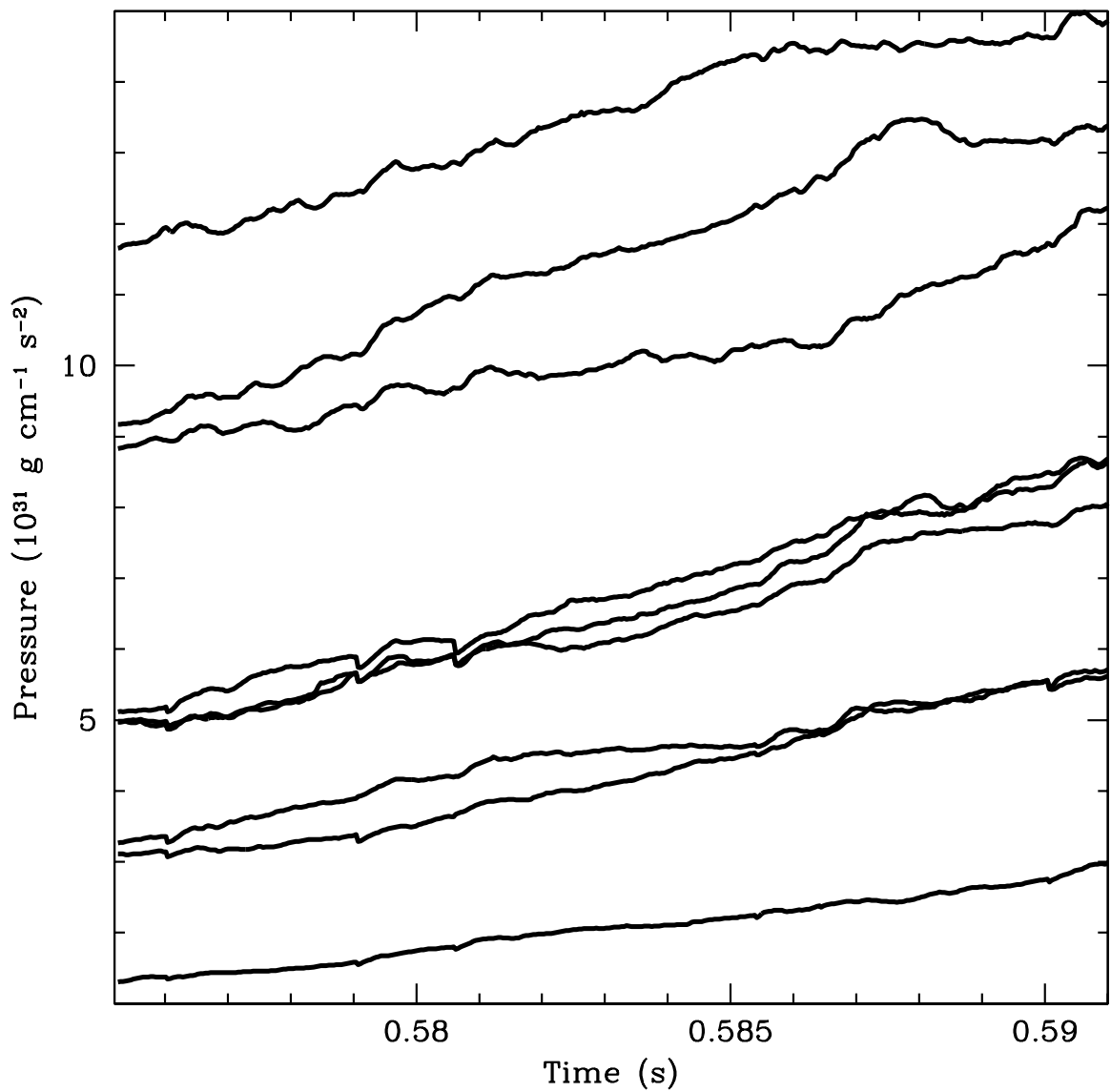}
\caption{Pressure versus time for 9 particles lying 35\,km from the 
center of the neutron star.  This plot has 400 time dumps corresponding 
to 0.04\,ms time resolution.  The primary evolution of the pressure 
is a steady increase as matter continues to pile onto the neutron star.  
Some variation exists on ms timescales, but at the 1\% level, corresponding 
to energy injection below $10^{44} {\rm \,erg \, s^{-1}}$.}
\label{fig:osci}
\end{figure}
\clearpage

\begin{figure}
\plotone{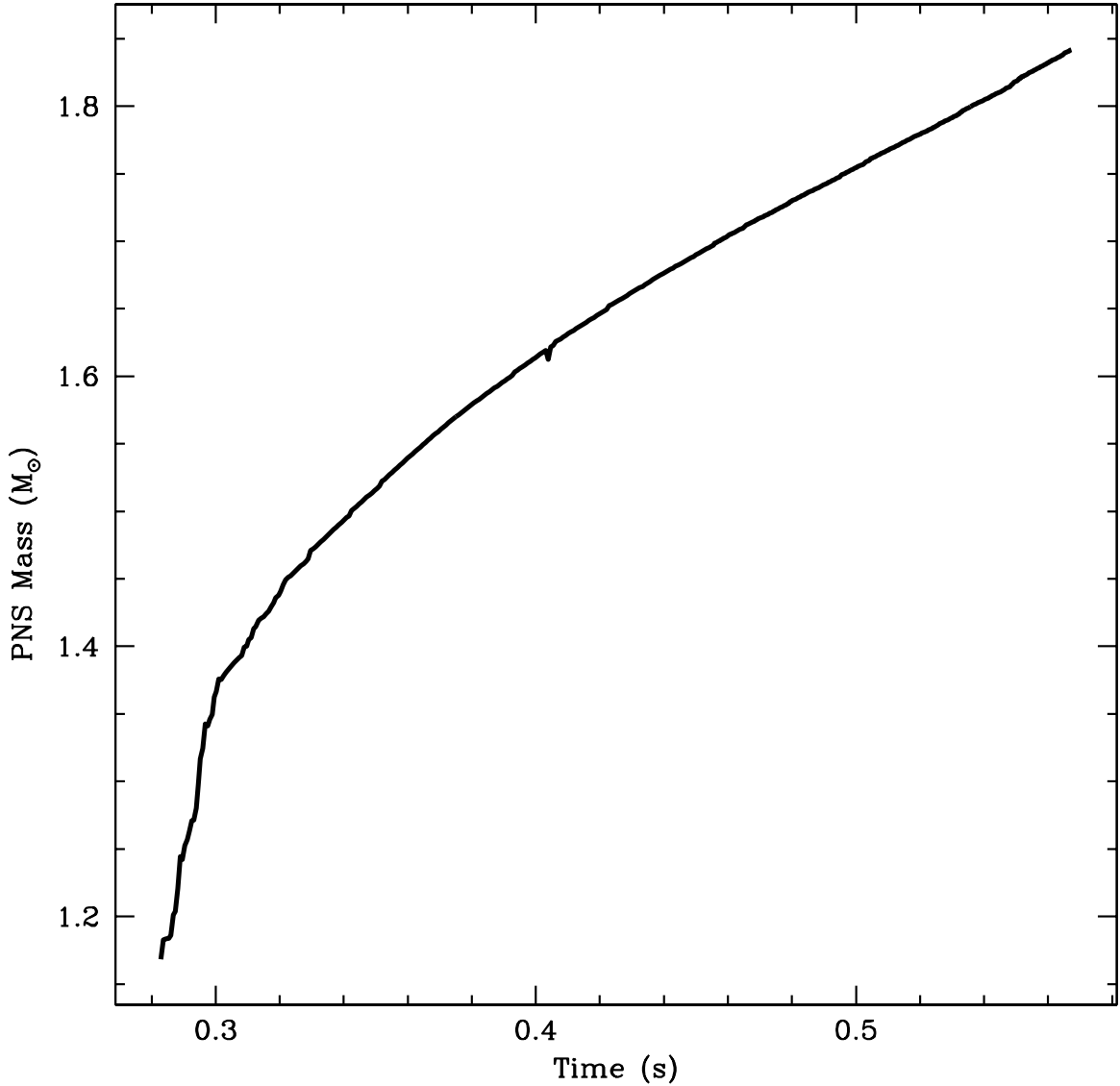}
\caption{Mass of the neutron star as a function of time.  Because 
of the long delay in this explosion, the mass of the neutron star 
has grown beyond 1.8\,M$_\odot$ prior to the launch of the explosion.  
Clearly such a long delay will not make a ``typical'' neutron star.  
Indeed, with the weak explosion energies from this explosion, we 
expect to have considerable fallback and the final remnant of this 
object will be a black hole.  This points out a difficulty with 
models requiring long delays - for stars above $\sim 15\,$M$_\odot$, 
such long delays produce massive neutron stars or black holes.}
\label{fig:nsmass}
\end{figure}
\clearpage

\begin{figure}
\plotone{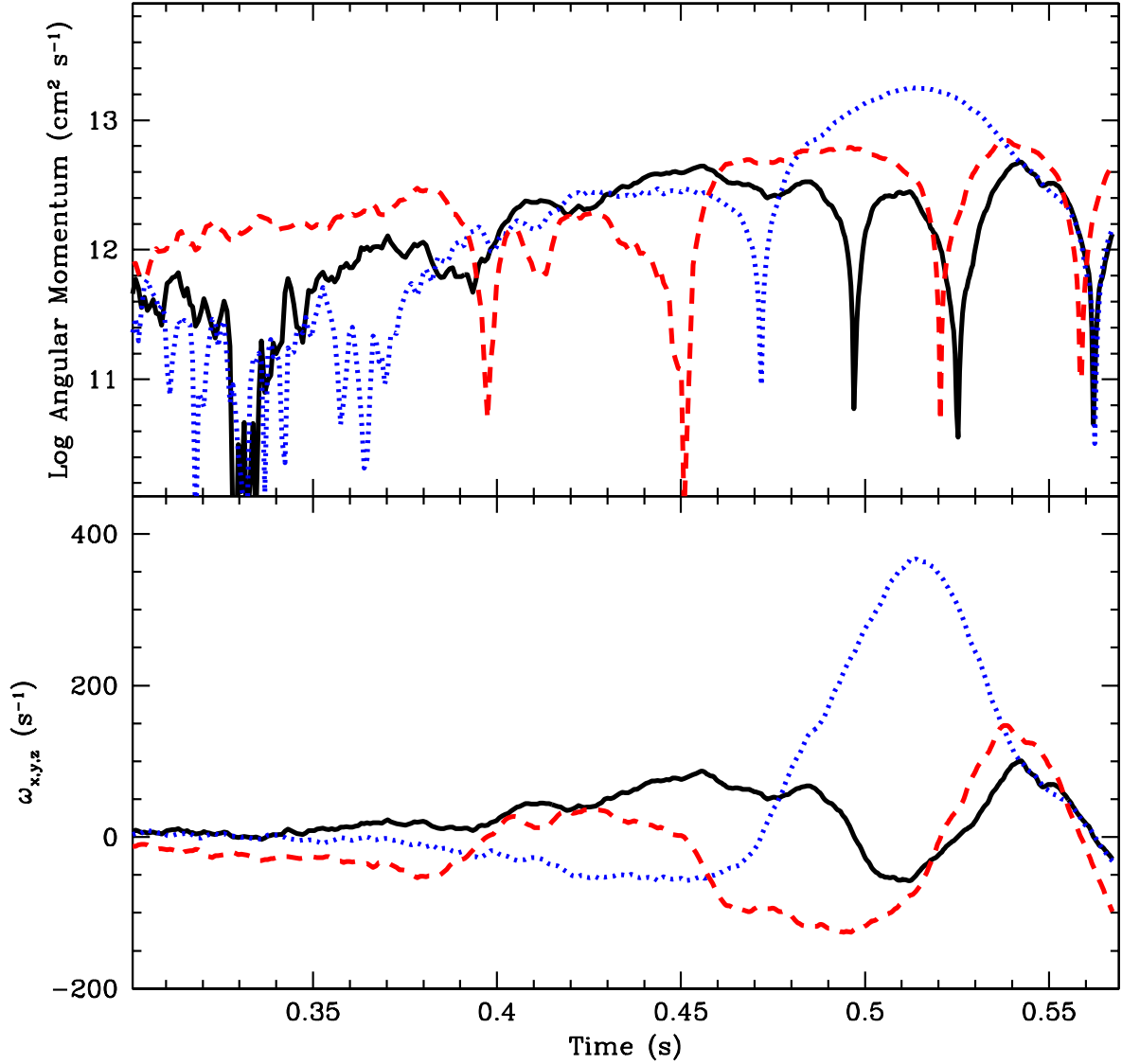}
\caption{The absolute value of the x (solid), y (dotted) and z
(dashed) angular momenta (top) of the neutron star as a function of
time.  The spikes correspond to a change in sign of the angular
momentum.  The bottom panel shows the corresponding x (solid), y
(dotted) and z (dashed) angular velocity as a function of time.  The
corresponding period is $2 \pi/ |\omega|$.  The neutron star achieves 
periods that are as fast as 16\,ms, but at the end of the simulation, 
the period is closer to 1\,s.}
\label{fig:nsspin}
\end{figure}
\clearpage

\begin{figure}
\plotone{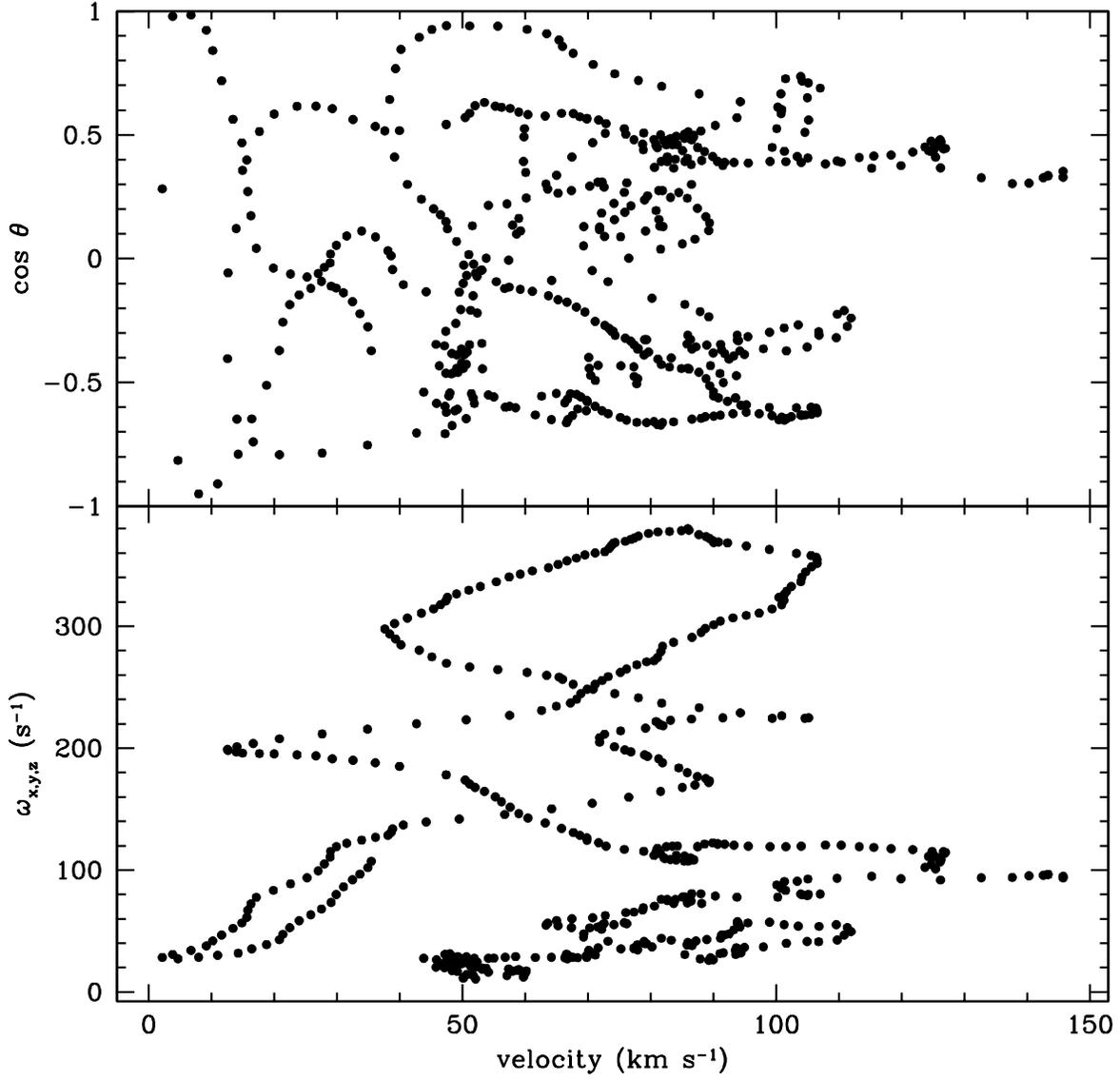}
\caption{The angle between neutron star velocity and spin vectors
(top) and magnitude of the spin velocity (bottom) as a function of the
magnitude of the velocity at different times in the explosion.  Although 
both the velocity and spin rate evolve through accretion, there is 
no correlation between the spin and the velocity.}
\label{fig:velspin}
\end{figure}
\clearpage

\begin{figure}
\plotone{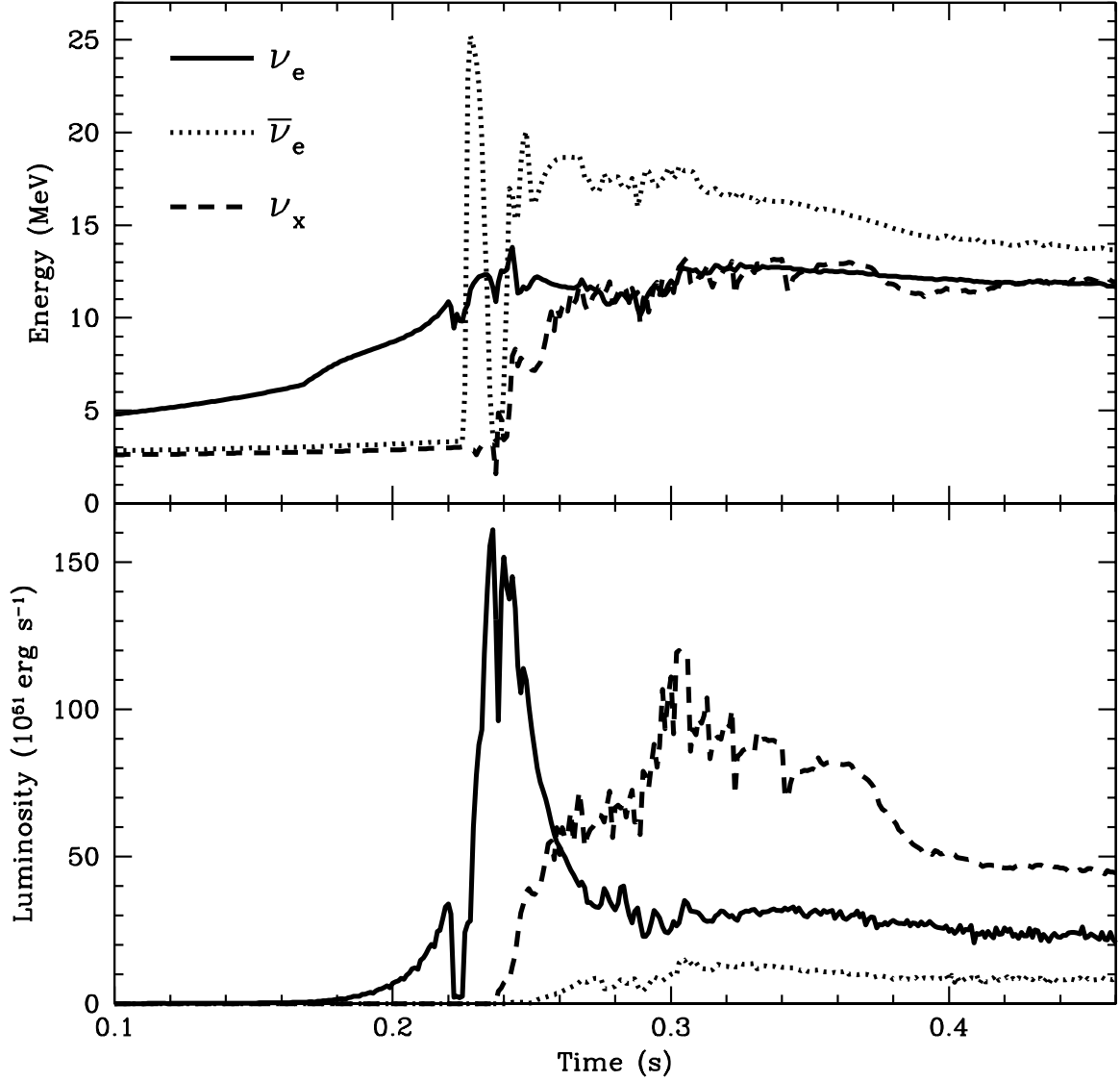}
\caption{Neutrino energy (top) and neutrino luminosity (bottom) as 
a function of time for the 3 species followed in the simulation:  
electron neutrino (solid), electron anti-neutrino (dotted), 
$\mu$ and $\tau$ neutrinos (dashed).  Note that the electron 
anti-neutrino energies are $<20\%$ higher than the electron 
neutrino energies at the end of the simulation, but their 
fluxes are a factor of 2 lower.}
\label{fig:neut}
\end{figure}
\clearpage

\begin{figure}
\plotone{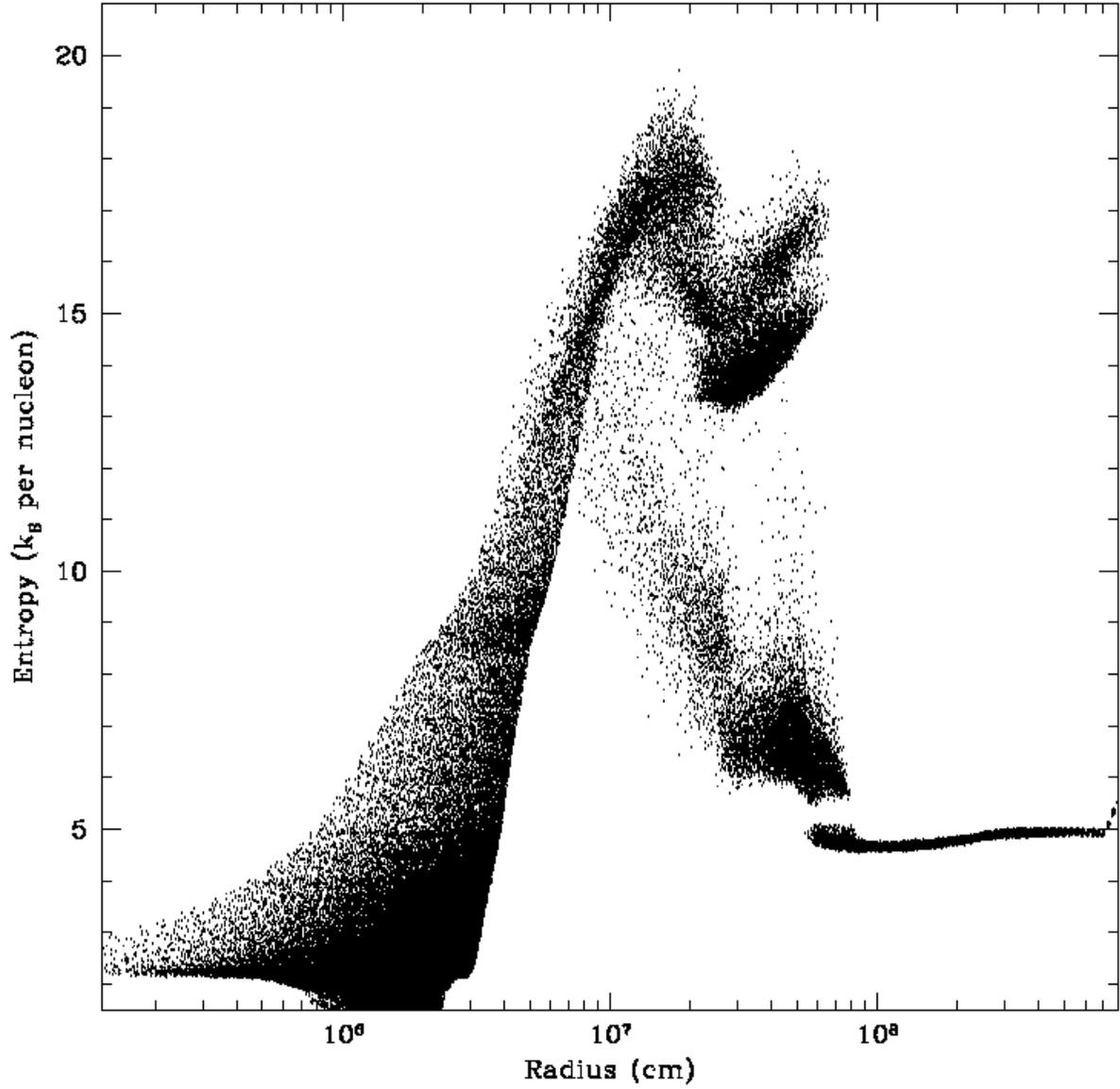}
\caption{Entropy (in units of Boltzmann constant per nucleon) as a
function of radius for our stellar core 600\,ms after collapse.  
Even at these late times, the peak entropy is less than 20\,$k_{\rm B}$ 
per nucleon.  The downflows are characterized by the low entropy 
material.}
\label{fig:entr}
\end{figure}
\clearpage

\end{document}